\documentclass[twocolumn,superscriptaddress,amsmath,amssymb,aps,prl]{revtex4-1}

\usepackage{graphicx}
\usepackage{dcolumn}
\usepackage{bm}

\usepackage[colorlinks,urlcolor=blue,citecolor=blue,linkcolor=blue]{hyperref}

\usepackage{physics}
\usepackage{comment}
\usepackage{color}

\newcommand{\up}{\uparrow}
\newcommand{\down}{\downarrow}
\renewcommand{\k}{{\bf k}}

\newcommand{\p}{{\bf p}}

\newcommand{\q}{{\bf q}}

\newcommand{\ek}{\epsilon_{\k}}
\newcommand{\ep}{\epsilon_{\p}}

\newcommand{\nn}{\nonumber}
\newcommand{\beq}{\begin{equation}}
\newcommand{\eeq}{\end{equation}}

\newcommand{\ri}{{\rm int}}

\newcommand{\ks}{{\k\sigma}}
\newcommand{\e}{{\text{ej}}}
\newcommand{\ii}{{\text{inj}}}

\newcommand{\hH}{\hat{H}}
\newcommand{\hc}{\hat{c}}

\newcommand{\hU}{\hat{U}}

\newcommand{\kB}{{k_B}}

\newcommand{\m}{{\rm med}}

\newcommand{\eb}{\epsilon_b}

\newcommand{\weizhe}[1]{{\color{black}#1}}

\begin{document}

\title{Radio-frequency response and contact of impurities in a quantum gas}

\author{Weizhe Edward Liu}
\affiliation{School of Physics and Astronomy, Monash University, Victoria 3800, Australia}
\affiliation{ARC Centre of Excellence in Future Low-Energy Electronics Technologies, Monash University, Victoria 3800, Australia}
\author{Zhe-Yu Shi}
\affiliation{School of Physics and Astronomy, Monash University, Victoria 3800, Australia}
\affiliation{State Key Laboratory of Precision Spectroscopy, East China Normal University, Shanghai 200062, China}
\author{Jesper Levinsen}%
\affiliation{School of Physics and Astronomy, Monash University, Victoria 3800, Australia}
\affiliation{ARC Centre of Excellence in Future Low-Energy Electronics Technologies, Monash University, Victoria 3800, Australia}
\author{Meera M.~Parish}
\affiliation{School of Physics and Astronomy, Monash University, Victoria 3800, Australia}
\affiliation{ARC Centre of Excellence in Future Low-Energy Electronics Technologies, Monash University, Victoria 3800, Australia}

\date{\today}

\begin{abstract}
We investigate the radio-frequency spectroscopy of impurities interacting with a quantum gas at finite temperature. In the limit of a single impurity, we show \weizhe{using Fermi's golden rule} that introducing (or injecting) an impurity into the medium is equivalent to ejecting an impurity that is initially interacting with the medium, since the ``injection'' and ``ejection'' spectral responses are simply related to each other by an exponential function of frequency. Thus, the full spectral information for the quantum impurity is contained in the injection spectral response, which can be determined using a range of theoretical methods, including variational approaches. We use this property to compute the finite-temperature equation of state and Tan contact of the Fermi polaron. Our results for the contact of a mobile impurity are in excellent agreement with recent experiments and we find that the finite-temperature behavior is qualitatively different compared to the case of infinite impurity mass.
\end{abstract}

\maketitle

The canonical problem of an impurity immersed in a quantum medium has attracted much attention recently due to the range of quantum-impurity scenarios that can be investigated using ultracold atomic gases. By varying the statistics of the atoms, it is possible to create an impurity in either a bosonic or fermionic medium, corresponding to a \textit{Bose polaron}~\cite{Catani2012,Hu2016,Jorgensen2016,Camargo2018,Yan2019nature} or a \textit{Fermi polaron}~\cite{Schirotzek2009,Nascimbene2009,Kohstall2012,Koschorreck2012,Zhang2012,Cetina2015,Cetina2016,Scazza2017,Oppong2019,Yan2019}, respectively. The behavior of the Fermi polaron provides fundamental insight into a variety of fermionic systems including spin-imbalanced Fermi gases~\cite{Chevy2006,Lobo2006,Navon2010}, excitons in doped semiconductors~\cite{Sidler2017},  
and protons in neutron stars~\cite{Kutschera1993}. The Bose polaron, on the other hand, is potentially relevant to systems involving particles coupled to a bosonic field, such as electrons coupled to phonons~\cite{Frohlich1954}, and quasiparticles in the vicinity of a phase transition.

\begin{figure}[hbt]
    \centering
    \includegraphics[width=0.9\columnwidth]{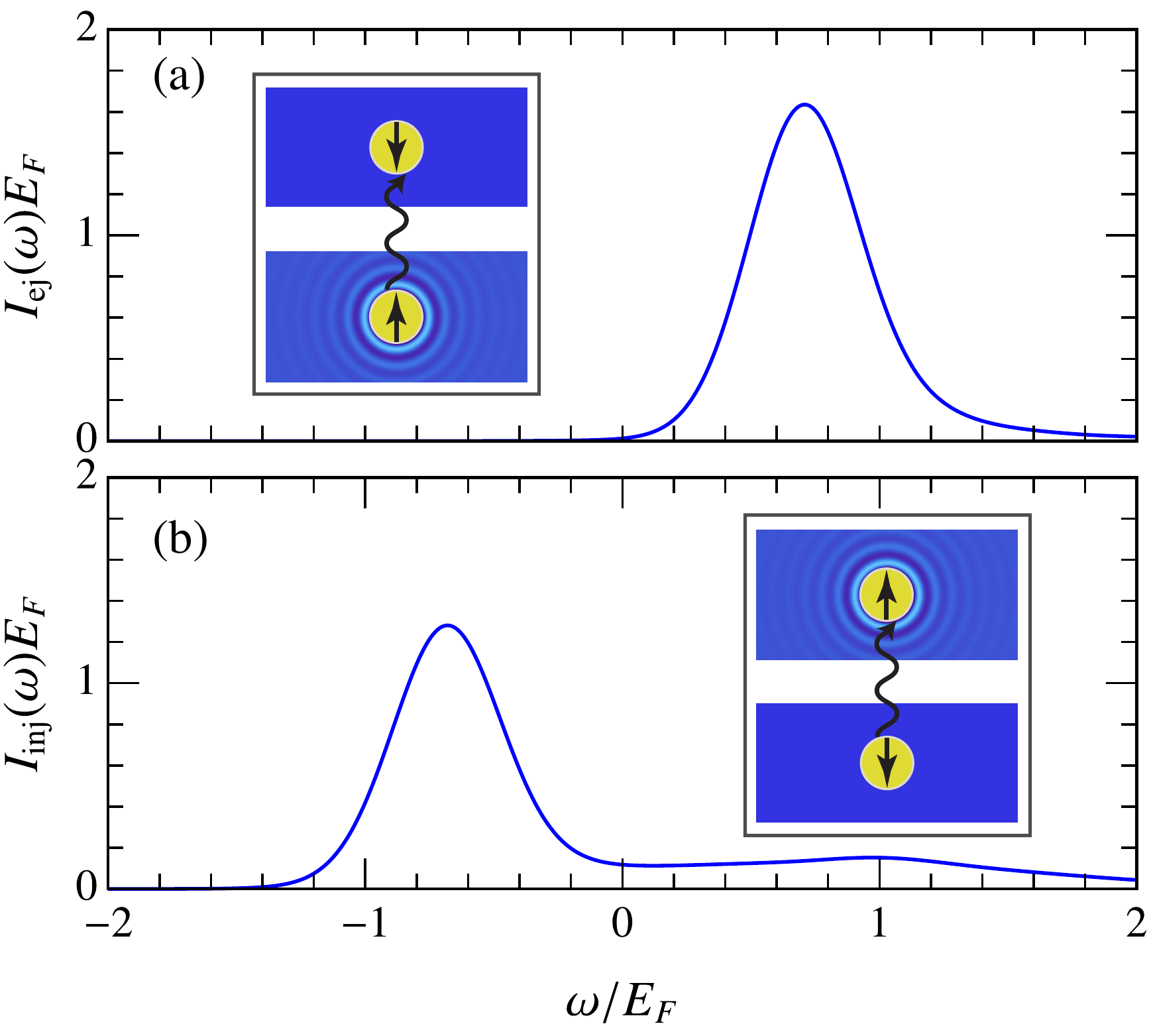}
    \caption{Examples of impurity ejection (a) and injection (b) spectra observable in a quantum-gas experiment. 
    An rf field of frequency $\omega$ drives a transition between the interacting ($\up$) and non-interacting ($\down$) impurity spin states (insets). We take the Fermi polaron at unitarity with mass $M=m$ and temperature $T=0.2T_F$.}
    \label{fig:rf}
\end{figure}

A key experimental probe of many-body physics in the cold-atom system is radio-frequency (rf) spectroscopy, whereby an applied rf pulse transfers atoms in one hyperfine spin state to another unoccupied state~\cite{Torma2014}. Two different spectroscopic protocols have been used to characterize the polaron \weizhe{in the linear response regime}, as depicted in Fig.~\ref{fig:rf}. In the standard ``ejection'' rf spectroscopy [Fig.~\ref{fig:rf}(a)], impurities are initially prepared in a spin state that interacts with the medium and are then driven into a non-interacting spin state by the rf field~\cite{Schirotzek2009,Koschorreck2012,Yan2019}. The rate of atoms ejected from the interacting state provides information about polaron properties such as the quasiparticle energy and lifetime~\cite{Punk2007,Schmidt2018,Mulkerin2019,Tajima2019}. In the opposite scheme in Fig.~\ref{fig:rf}(b), one instead measures the rate of impurities that are transferred from the non-interacting spin state to the interacting state~\cite{Kohstall2012,Cetina2015,Cetina2016,Scazza2017,Oppong2019}. This ``injection'' spectral response has the advantage that it is easier to calculate theoretically~\cite{Massignan2008,Schmidt2011,Rath2013,Li2014,Massignan2014,Li2014,Parish2016,Jorgensen2016,Shchadilova2016,Goulko2016,Sun2017,Guenther2018,Hu2018,Liu2019,Tajima2019,Dzsotjan2019u,Field2020} and it can be measured in less stable atomic gases where it is difficult to achieve thermal equilibrium of the interacting system. However, such a spectral response appears to lack important thermodynamic information such as the Tan contact~\cite{Tan2008}, which is only expected to be visible in ejection spectra~\cite{Braaten2010}.

In this Letter, we \weizhe{use Fermi's golden rule to} show that the description of injection and ejection spectra at a given temperature $T$ is greatly simplified in the single-impurity limit. Crucially, we find that the ejection rate $I_\e$ and injection rate $I_\ii$ of the impurity 
obey the fundamental relationship
\begin{align} \label{eq:ej-inj}
    I_\e(\omega) = e^{\beta \hbar\omega} e^{\beta \Delta F} I_\ii(-\omega), 
\end{align}
where $\omega$ is the frequency of the rf field, $\beta = 1/(k_BT)$, and $\Delta F \equiv F - F_0$ is the difference in free energies of the system with and without impurity-medium interactions. Note that this relationship is independent of dimensionality or the type of medium; it only requires the system to initially be in thermal equilibrium.
\weizhe{We also assume that the fraction of impurities transferred to the final state is sufficiently small so that linear response theory (and therefore Fermi's golden rule) is valid. Such a regime can be routinely achieved in cold-atom experiments (see, e.g., Ref.~\cite{Cetina2016}).} 

Equation~\eqref{eq:ej-inj} allows us to obtain the ejection spectrum as shown in Fig.~\ref{fig:rf} from the impurity injection spectrum. The latter can be determined using methods such as our recently developed finite-temperature variational principle~\cite{Liu2019}. Focussing on the case of the Fermi polaron, we use Eq.~\eqref{eq:ej-inj} to compute the finite-temperature equation of state $\Delta F$ for the first time, and we find excellent agreement with recent experimental measurements of the Tan contact~\cite{Yan2019}. We furthermore obtain the exact equation of state when the impurity mass is infinite and we find that the temperature dependence is qualitatively different from that of a mobile impurity.

\paragraph{Impurity rf response ---} 
In the following, we consider an impurity immersed in a quantum medium at non-zero temperature $T$. We assume that the impurity can exist in two different spin states, where $\down$ is non-interacting with the medium and $\up$ can have arbitrarily strong interactions with the surrounding medium. We can write the general Hamiltonian as:
\begin{align}
    \hH_\ri&= \hH_0 +\hU,
\end{align}
where $\hU$ describes the tunable interactions between spin-$\up$ impurities and the medium, while the non-interacting part of the Hamiltonian is
\begin{align}
\hH_0=\sum_\ks \epsilon_\k \hc_\ks^\dagger \hc_\ks + \hH_\m.
\end{align}
Here $\hc_\ks^\dagger$ creates an impurity with spin $\sigma$, momentum $\k$ and kinetic energy $\epsilon_\k = |\k|^2/(2M) \equiv k^2/(2M)$, with $M$ the impurity mass. Note that we set $\hbar$ and $\kB$ to 1 here and in the following. We initially consider the medium-only Hamiltonian $\hH_\m$ to be general, since our arguments do not depend on the precise details of the medium.

We start with a single impurity that is initially interacting or non-interacting with the medium, corresponding to the impurity spin $\up$ or $\down$, respectively. We then apply a weak rf field of frequency $\omega$ that drives the impurity into the opposite spin state (see Fig.~\ref{fig:rf}). Using Fermi's golden rule, we can obtain the ejection and injection transition rates, respectively, for a given momentum $\p$ of the non-interacting impurity~\cite{Liu2020_long}:
\begin{subequations}
\label{eq:spec}
\begin{align}
    \label{eq:spec_ej}
    A_\e(\p,\omega)&
    \!=\sum_{n,\nu}\frac{e^{-\beta E_{\nu}}}{Z_\ri}|\!\bra{n}\hc_{\p\up}\ket{\nu}\!|^2 \delta(\omega+ E_{\nu n\p}),\\
    A_\ii(\p,\omega)& 
    \!=\sum_{n,\nu}\frac{e^{-\beta E_n}}{Z_\m}|
    \!\bra{n}\hc_{\p\up}\ket{\nu}\!|^2
    \delta(\omega - E_{\nu n\p}),
\end{align}
\end{subequations}
with $E_{\nu n\p} \equiv E_\nu - E_n -\epsilon_\p$. The eigenstates and associated energies of the total interacting system are represented by $\ket{\nu}$ and $E_{\nu}$, respectively, while $\ket{n}$ and $E_{n}$ correspond to the medium-only eigenstates and energies, respectively. We have also defined the partition functions $Z_\ri = \sum_\nu e^{-\beta E_\nu}$ and $Z_{\m} = \sum_n e^{-\beta E_n}$. Note that the applied rf field is momentum conserving on the scale of the cold-atom cloud~\cite{Haussmann2009}, and thus we can take the non-interacting impurity to have a well-defined momentum in the final (initial) state in the case of ejection (injection) spectroscopy. Indeed, the ejection spectral function in Eq.~\eqref{eq:spec_ej} has been measured for Fermi polarons in two dimensions using momentum-resolved spectroscopy~\cite{Koschorreck2012}. Note, further, that we have neglected an overall prefactor that corresponds to the strength of the rf field.

Thus far, we have made no assumptions about the medium other than that it is initially in a thermal state (either with or without the impurity). Using the properties of delta functions, it is straightforward to obtain from Eq.~\eqref{eq:spec} the general relation
\begin{align} \label{eq:dbal}
    A_\e(\p,\omega)=
    \frac{Z_\m}{Z_\ri}e^{\beta\omega} e^{-\beta\epsilon_\p}
    A_\ii(\p,-\omega),
\end{align}
which is intimately connected to the detailed balance condition in thermal equilibrium~\cite{Liu2020_long}. Such a relation can in principle be generalized to other types of impurities, e.g., immobile impurities with an internal degree of freedom.

The rf spectroscopy employed in experiment is typically averaged over all impurity momenta, and thus 
we consider the total transition rates
\begin{align} \nn
    I_\e(\omega)= \sum_\p A_\e(\p,\omega), \quad   
    I_\ii(\omega) = 
    \sum_\p \frac{e^{-\beta\epsilon_\p}}{Z_{\rm imp}} 
    A_\ii(\p,\omega) .
\end{align}
For the case of injection, we have assumed that the initially non-interacting impurity is in thermal equilibrium with the medium and thus its momentum satisfies the Boltzmann distribution, with impurity partition function $Z_{\rm imp} = \sum_\k e^{-\beta \epsilon_\k}$. Combining this with Eq.~\eqref{eq:dbal} then yields the key result in Eq.~\eqref{eq:ej-inj}, once we identify the free energy difference
\begin{align}
    \Delta F \equiv F - F_0 = T\ln\left(\frac{Z_\m Z_{\rm imp}}{Z_\ri}\right) .
\end{align}
As we show below, one can extract $\Delta F$ from  Eq.~\eqref{eq:ej-inj} by applying the sum rules for the spectral functions.

\paragraph{Fermi polaron ---}
We now turn to the well-studied example of the Fermi polaron, where the medium is a non-interacting spinless Fermi gas in three dimensions:
\begin{align}
    \hH_\m = \sum_\k \left(\ek^\m -\mu \right) \hat{f}_\k^\dag \hat{f}_\k .
\end{align}
Here $\mu$ is the chemical potential of the Fermi gas, $\ek^\m = k^2/(2m)$, and $\hat{f}^\dag_\k$ creates a fermionic atom with momentum $\k$ and mass $m$. We choose $\mu$ such that the density of the Fermi gas is fixed, with Fermi momentum $k_F$ and corresponding Fermi energy $E_F = k_F^2/(2m)$.

The impurity is assumed to have short-range attractive interactions with the Fermi gas, which is described by~\footnote{Here we assume that the impurity-medium interactions can be described by a single-channel Hamiltonian, which corresponds to a broad Feshbach resonance. We consider the more general two-channel model in Ref.~\cite{Liu2020_long}.}
\begin{align}
    \hU = \frac{u}{V} \sum_{\k,\k',\q} \hat{f}_\k^\dag \hat{f}_{\k-\q} \,
    \hat{c}_{\k'\up}^\dag \hat{c}_{\k'+\q, \up}  ,
\end{align}
where the interaction strength $u$ is related to the $s$-wave scattering length $a$ via  $\frac{1}{u} = \frac{m_r}{2\pi a}-\frac{1}{V} \sum_\k \frac{2m_r}{k^2}$, with $V$ the system volume, and reduced mass $m_r = mM/(m + M)$. We focus on equal masses, $m=M$, which corresponds to the case in recent experiments~\cite{Scazza2017,Yan2019}.

We determine the impurity spectral response using the finite-temperature variational principle for injection spectroscopy~\cite{Liu2019}, where one describes the time evolution of the injected impurity with the approximate operator
\begin{align}
    \hc_{\p \up}(t) \simeq \alpha_{\p;0}(t) \hc_{\p \up} + \sum_{\k \ne \q} \alpha_{\p;\k\q}(t) \hat{f}_{\q}^\dagger \hat{f}_{\k} \,
    \hc_{\p-\k+\q,\up} ,
\end{align}
which includes up to one particle-hole excitation of the Fermi medium. Here the $\alpha$'s are complex time-dependent variational parameters. By minimizing the error in the impurity time evolution~\cite{Liu2019,Liu2020_long}, one obtains a set of coupled linear equations for the variational parameters that yields the stationary solutions $\alpha_{\p;0}^{(l)} e^{-i E_\p^{(l)}t}$ and $\alpha_{\p;\k\q}^{(l)} e^{-i E_\p^{(l)}t}$, with energy eigenvalue $E_\p^{(l)}$. This then allows us to compute the injection spectral function for the Fermi polaron~\cite{Liu2019,Liu2020_long}
\begin{align} \label{eq:Ainj}
    A_\ii(\p,\omega) = \sum_l \abs{\alpha_{\p;0}^{(l)}}^2 \delta(\omega + \ep -E_{\p}^{(l)}) .
\end{align}

In Fig.~\ref{fig:rf}(b) we show the momentum-averaged injection spectrum $I_\ii$ at unitarity $1/a = 0$ and finite temperature $T = 0.2 T_F$, where we have Fermi temperature $T_F \equiv E_F$. We clearly observe an attractive polaron peak at $\omega \simeq - 0.7 E_F$ and a broad spectral feature at positive frequencies. Using Eq.~\eqref{eq:ej-inj}, we also obtain the ejection spectrum $I_\e$, and we see in Fig.~\ref{fig:rf}(a) that the broad feature is suppressed and only the polaron peak is visible, consistent with experiment~\cite{Yan2019} and with previous theoretical calculations for a finite density of impurities~\cite{Mulkerin2019,Tajima2019}. Note, further, that $I_\e$ exhibits a power-law tail at large positive frequencies which is related to the contact $C$~\cite{Braaten2010}, defined via the thermodynamic relation~\cite{Tan2008}
\begin{align} \label{eq:contact}
    C & = 8\pi m_r \left. \pdv{F}{(-1/a)}\right|_{T,\mu} = 8\pi m_r \left. \pdv{\Delta F}{(-1/a)} \right|_{T,\mu} .
\end{align}
\weizhe{For clarity, we have convoluted the discrete spectra obtained from our variational approach with a Gaussian of width $0.2 E_F$~\cite{Parish2016,Liu2019,Liu2020_long}}.

We now extract the difference in free energies directly from the 
spectral response. Since we are working with a single impurity whose internal state is unchanged by the interactions with the medium, we have the sum rule $\int d\omega \, I_\e(\omega) = 1$. Thus, integrating Eq.~\eqref{eq:ej-inj} over frequency and using Eq.~\eqref{eq:Ainj}, we finally obtain
\begin{align}
    \Delta F = -T \ln \left[\frac{\sum_{\p,l} e^{-\beta E_{\p}^{(l)}} \abs*{\alpha_{\p;0}^{(l)}}^2}{\sum_\p e^{-\beta \ep}} \right] .  
\end{align}

\begin{figure}[bht]
    \centering
    \includegraphics[width=0.95\columnwidth]{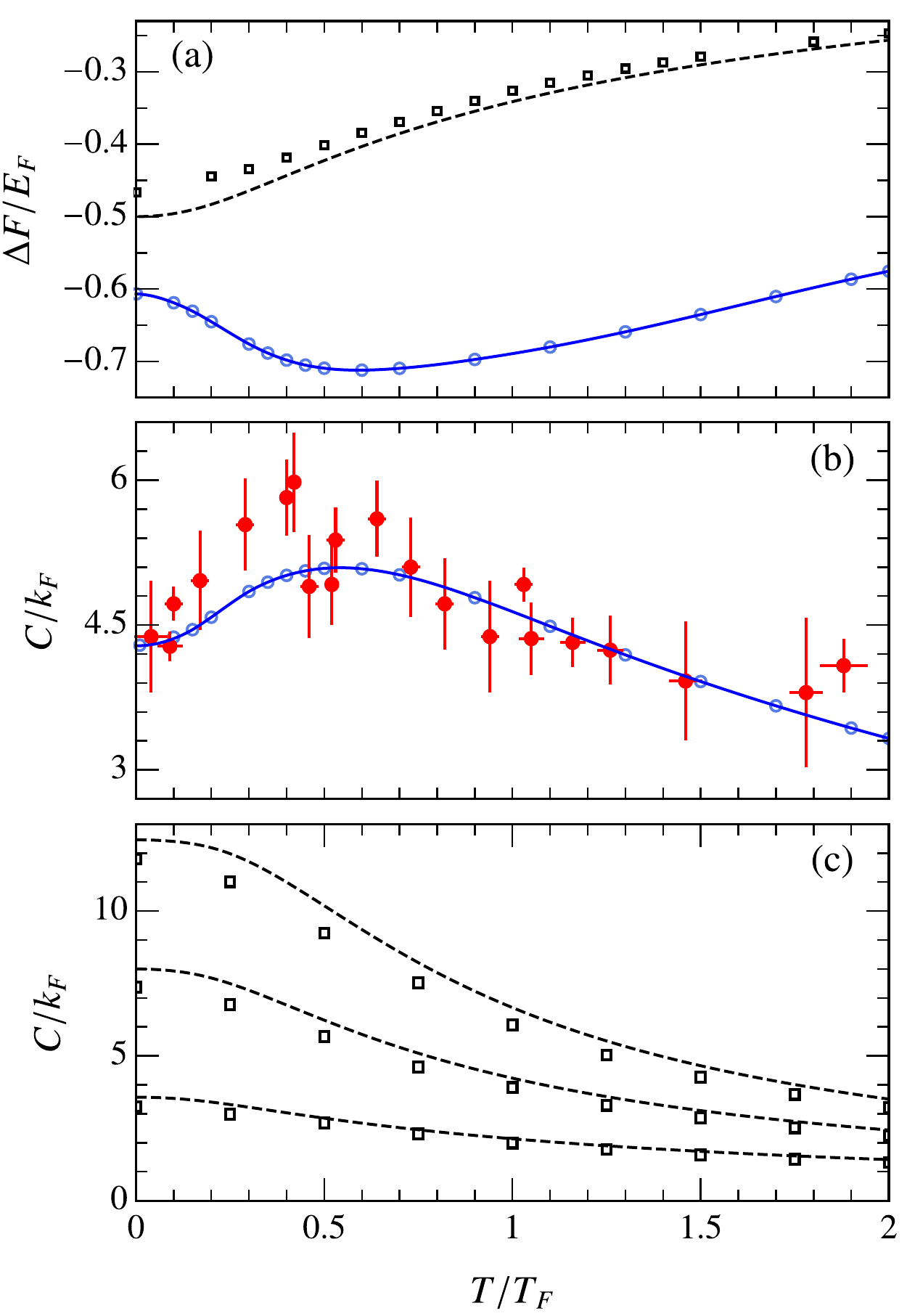}
    \caption{Finite-temperature equation of state for the Fermi polaron. (a) Relative free energy $\Delta F$ at unitarity for both equal-mass (blue solid) and infinite-mass cases \weizhe{(black dashed, exact result, Eq.~\eqref{eq:DeltaFinf}; black squares, variational approach)}. (b) Tan contact at unitarity for the equal-mass case obtained in this work (blue solid) and from experiment in Ref.~\cite{Yan2019} (red filled circles). (c) Tan contact for the infinite-mass case \weizhe{(black dashed, exact result, Eq.~\eqref{eq:con_inf}; black squares, variational)} at interaction strengths $1/(k_F a) = -0.5, 0, 0.3$ (bottom to top).}
    \label{fig:EoS}
\end{figure}

Our results for $\Delta F$ as a function of temperature are displayed in Fig.~\ref{fig:EoS}(a), where we once again consider the unitarity limit $1/a = 0$. At $T=0$ we recover the ground-state polaron energy $-0.61 E_F$ expected from the Chevy ansatz~\cite{Chevy2006}\weizhe{, which agrees well with state-of-the-art quantum Monte Carlo~\cite{Houcke2020}}. However, when we increase the temperature from zero, we observe a striking non-monotonic behavior, where $\Delta F$ shifts to more negative values for $T\lesssim 0.5 T_F$ before tending towards zero at higher temperatures. A similar behavior is also apparent in the calculated position of the polaron peak in both injection and ejection spectra~\cite{Hu2018,Mulkerin2019,Tajima2019,Liu2020_long}.

In Fig.~\ref{fig:EoS}(b), we see that the non-monotonic dependence on temperature is also mirrored in the contact, which we determine using Eq.~\eqref{eq:contact}. In particular, our results are in excellent agreement with recent experiments on the equal-mass Fermi polaron at unitarity~\cite{Yan2019}. Thus, the impurity-medium interactions appear to have the strongest effect on the system when $T \simeq 0.5 T_F$. Note that we must always have $\pdv{C}{T} = 0$ at $T = 0$ since the entropy of the system $\pdv{\Delta F}{T}$ tends to zero in this limit.

To gain further insight into this behavior, we consider the limit of infinite impurity mass $M \to \infty$, which can be reduced to the problem of a single particle in a static potential and is thus exactly solvable~\cite{Schmidt2018}. In this case, the contact in Eq.~\eqref{eq:contact} becomes~\cite{Liu2020_long}
\begin{align}\label{eq:con_inf}
    C & = \frac{16\pi^2}{V} \sum_\k \frac{a^2}{1+k^2a^2}
    n_F(\epsilon_\k^\m) + \theta(a) \frac{8\pi}{a} n_F(\eb),
\end{align}
with Fermi-Dirac distribution $n_F(\epsilon) = \left(1+e^{\beta(\epsilon-\mu)}\right)^{-1}$. Here we have simply thermally averaged the contact over all the single-particle eigenstates in the fixed impurity potential. These consist of scattering states of well-defined momentum $\k$, and a bound state when $a>0$ with energy $\eb = -1/(2ma^2)$.

As shown in Fig.~\ref{fig:EoS}(c), the infinite-mass contact from Eq.~\eqref{eq:con_inf} always monotonically decreases with increasing temperature, in contrast to the equal-mass case. This monotonic dependence is due to the fact that the contact of each single-particle eigenstate decreases with increasing eigenstate energy, as we can see in Eq.~\eqref{eq:con_inf}. Such behavior is also present in the relative free energy $\Delta F$, which can be obtained from Eq.~\eqref{eq:contact} by integrating Eq.~\eqref{eq:con_inf} over the scattering length $a$. At unitarity, we have the simple universal expression
\begin{align}
    \Delta F = -\frac{T}{2} \ln\left[1+ e^{\beta\mu} \right],
    \label{eq:DeltaFinf}
\end{align}
which is plotted in Fig.~\ref{fig:EoS}(a). \weizhe{For comparison, we have also included the results from the variational approach, which we see lie close to the exact curves.}

The results for the infinitely massive impurity imply that impurity recoil is necessary to produce the non-monotonic behavior observed in Fig.~\ref{fig:EoS} and in experiment~\cite{Yan2019}. In particular, the presence of recoil can constrain the scattering of the impurity with a fermion from the medium, thus resulting in an admixture of higher angular momentum channels which lowers the contact of the ground-state polaron (since the contact is zero outside of the $s$-wave channel). This in turn can lead to many-body states with a larger contact than in the polaron ground state, thus producing the non-monotonic temperature dependence in Fig.~\ref{fig:EoS}(b). A similar effect has also been observed for the contact of trapped few-body systems~\cite{Yan2013}.

To conclude, we have shown that there exists a simple mapping between the injection and ejection spectra of a quantum impurity. We have applied this to the case of the Fermi polaron and obtained the finite-temperature equation of state for the first time. We have also derived exact results for the equation of state in the limit of infinite impurity mass, thus shedding light on recent experimental measurements of the contact~\cite{Yan2019}. Our results pave the way for further studies of the spectral and thermodynamic properties of a host of other quantum impurity problems.
\weizhe{An interesting future direction is the regime beyond linear response, which is yet to be properly understood in quantum gases~\cite{Kohstall2012}, but which could in principle be investigated with our finite-temperature variational approach~\cite{Adlong2020}.}

\acknowledgements 
We are grateful to M.~W.~Zwierlein and H.~S.~Adlong for useful discussions, and we thank Z.~Yan for providing us with the experimental data from Ref.~\cite{Yan2019}.  JL and MMP acknowledge support from the Australian Research Council Centre of Excellence in Future Low-Energy Electronics Technologies (CE170100039). JL is also supported through the Australian Research Council Future Fellowship FT160100244. J. L., Z. Y. S., and M. M. P. acknowledge support from the Australian Research Council via Discovery Project No. DP160102739.

\end{document}